\documentclass[10pt]{iopart}
\usepackage{graphicx}
\usepackage{SIunits}

\begin{document}

\title[Superconducting coplanar microwave resonators up to 50\,GHz]{Superconducting coplanar microwave resonators with operating frequencies up to 50\,GHz}

\author{Desir\'{e}e~S~Rausch$^1$, Markus~Thiemann$^1$, Martin~Dressel$^1$, Daniel~Bothner$^2$\footnote{Present address:
Kavli Institute of NanoScience, Delft University of Technology, PO Box 5046, 2600 GA, Delft, The Netherlands.}, Dieter~Koelle$^2$, Reinhold~Kleiner$^2$, Marc~Scheffler $^1$}
\address{$^1$ 1.~Physikalisches Institut, Universit\"{a}t Stuttgart, Pfaffenwaldring 57, 70569 Stuttgart, Germany}
\address{$^2$ Physikalisches Institut and Center for Quantum Science (CQ) in LISA$^+$, Universit\"{a}t T\"{u}bingen, Auf der Morgenstelle 14, 72076 T\"{u}bingen, Germany}
\ead{scheffl@pi1.physik.uni-stuttgart.de}

\vspace{10pt}
\begin{indented}
\item[]\today
\end{indented}

\begin{abstract}
We demonstrate the operation of superconducting coplanar microwave resonators in a very large frequency range up to 50\,GHz.
The resonators are fabricated from niobium thin films on sapphire substrates and optimized for these high frequencies by small chip sizes. We study numerous harmonics of the resonators at temperatures between 1.5\,K and 6\,K, and we determine quality factors of up to 25000 at 1.5\,K. As an example for spectroscopy applications of such resonators we detect the superconducting transition of a bulk tin sample at multiple probing frequencies. 
\end{abstract}

\ioptwocol

\section{Introduction}
Superconducting coplanar microwave resonators, i.e.\ sections of a superconducting coplanar waveguide of finite length that are capacitively or inductively coupled to external circuitry, can drastically surpass their metallic equivalents in terms of higher quality factors $Q$, but at the expense of cryogenic cooling \cite{Frunzio2005,Hammer2007,Goeppl2008,Javaheri2016}. Such superconducting coplanar resonators are used for highly sensitive detectors \cite{Day2003,Hammer2007,Barends2007,Zmuidzinas2012} and for chip-based quantum information science \cite{Goeppl2008,Barends2014}. Cryogenic microwave studies of material properties also employ planar superconducting resonators: the material under study either constitutes the resonator or is brought close to it \cite{DiIorio1988,Langley1991,Porch1995,Scheffler2013,Hafner2014,Ghigo2016}.
There are numerous ongoing efforts to improve the performance of such superconducting resonators, e.g.\ concerning higher $Q$ \cite{Megrant2012,Quintana2014,Goetz2016} or reduced susceptibility to magnetic fields \cite{Plourde2009,Bothner2012,Singh2014,Ghirri2015}.
Our present work addresses another fundamental aspect, namely the operating frequency of superconducting coplanar resonators, which typically is in the range 1-20\,GHz and which we extend to frequencies as high as 50\,GHz.

For room-temperature applications there are clear objectives for coplanar resonators at frequencies higher than 20\,GHz \cite{Thompson2004, Chuang2011}, and there are also perspectives for the cryogenic applications mentioned above: circuit quantum electrodynamics (circuit QED) at higher resonator frequency could profit from less stringent cooling \cite{Wallraff2004} and frequency-multiplexed detectors from larger bandwidth, while facing the downside that operating superconducting resonators at higher frequencies fundamentally restricts $Q$ to lower values due to the strongly frequency-dependent conductivity of superconductors \cite{Pracht2013}. But the main reasons why so far superconducting coplanar resonators were not reported for frequencies above 20\,GHz are more mundane and yet inhibiting. Commercial instruments and components such as amplifiers, circulators, or connectors for frequencies above 20\,GHz are much less available and more costly than their lower-frequency counterparts. Furthermore, attenuation in metal-based microwave transmission lines increases strongly with increasing frequency due to the reduced skin depth \cite{Gustrau2012, Kurpiers2017} as well as due to smaller geometrical cross sections that are required to avoid undesired transmission modes \cite{Moding}. Finally, standing waves in cryogenic microwave lines caused by partial reflections at discontinuities such as connectors become more difficult to handle due to the smaller wavelength. All these reasons contribute to the present situation that cryogenic microwave experimentation is well established and rapidly growing for low-GHz frequencies, but corresponding works for the 20-50\,GHz range are hardly found in the literature.
However, in spectroscopy these inconveniences have to be overcome if interesting features of materials under study lie in this particular spectral range, as is often the case for superconducting or correlated materials \cite{Pracht2013,Basov2005,Scheffler2005c,Maeda2005,Steinberg2008,Basov2011}.
Also recent developments in quantum information science based on defects in diamond tend towards higher GHz frequencies \cite{Aslam2015}, e.g.\ considering the SiV defect \cite{Neu2011,Rogers2014,Becker2018}.

Optical spectroscopy in the range 20\,GHz to 100\,GHz is extremely demanding as the frequencies are higher than for conventional microwave experiments \cite{Pompeo2007} and too low for far-field THz optics \cite{Pracht2013,Hering2007, Geiger2016}, but recently this spectral range has been attacked from the low-frequency side: cryogenic Corbino reflectometry has been extended in frequency \cite{Tosoratti2000,Scheffler2005c,Scheffler2007,Steinberg2008,Scheffler2010}, well beyond the established 20~GHz range \cite{Booth1994,Scheffler2005a}, and cryogenic broadband coplanar lines were demonstrated even up to 67~GHz \cite{Clauss2013,Wiemann2015}. 
Superconducting planar resonators with their high $Q$ enable much higher sensitivity than these broadband approaches, and when compared to traditional three-dimensional cavities, they are more compact (for operation in dilution refrigerators or magnets) and very flexible concerning choice of multiple operating frequencies \cite{Scheffler2015}.
Therefore, we develop and operate superconducting coplanar resonators with frequencies extending up to 50\,GHz.

\begin{figure*}
	\includegraphics[width=\textwidth]{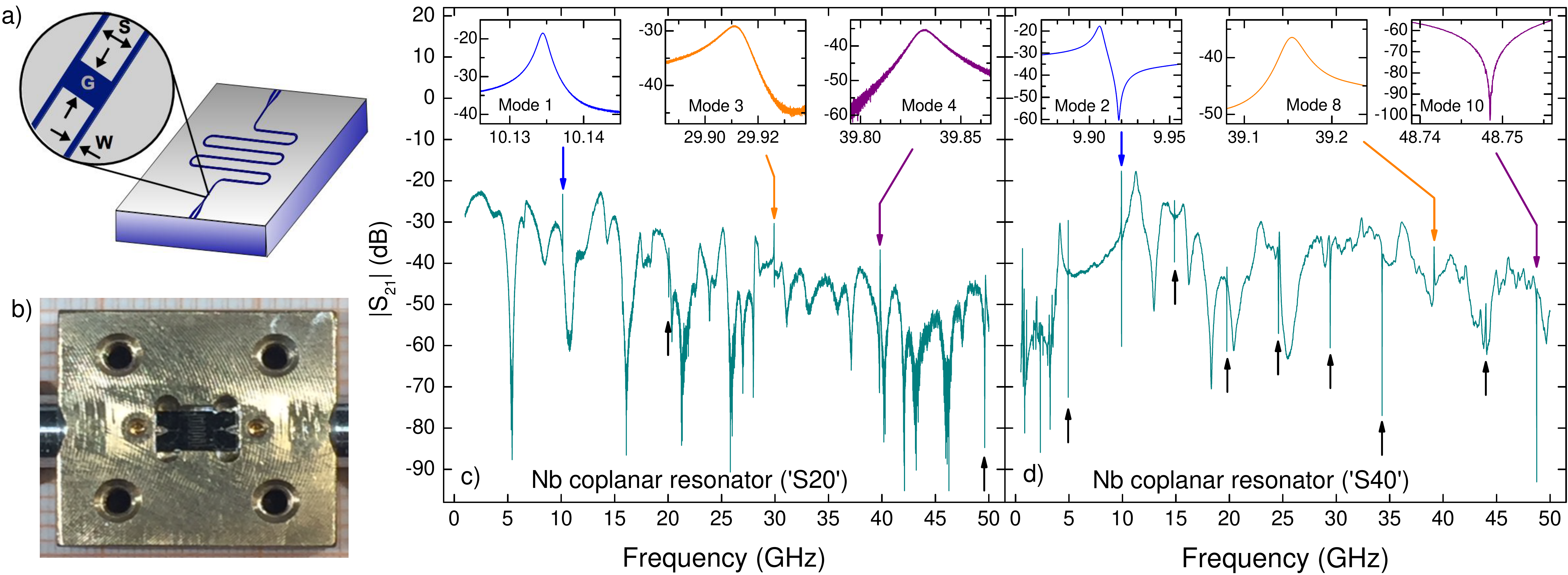}
	\caption{ Superconducting coplanar resonators for frequencies up to 50\,GHz. 
(a) Schematic view of the coplanar waveguide resonator design with characteristic geometric parameters. On top of a 430\,$\mu$m thick dielectric sapphire substrate (blue), a 200\,nm thick niobium layer is sputter-deposited. The meander line structure of the resonator is created by UV-lithography: this coplanar structure consists of a center conductor (strip width $S$) and two outer conductors that are separated by a distance $W$ from the center conductor. (b) Photograph of brass box with mounted resonator chip. Uncorrected transmission spectra of superconducting coplanar microwave resonators are shown in (c) for resonator \lq S20\rq{} with fundamental mode frequency of 10\,GHz and (d) for resonator \lq S40\rq{} with fundamental mode frequency of 5~GHz. Both spectra are taken at 1.8\,K in a frequency range up to 50~GHz. The arrows mark the modes. The insets show exemplary individual modes of the spectra.}
	\label{Spec}
\end{figure*}

\section{Experiment}
For this study we used coplanar resonators in a geometry as schematically shown in Fig.\ \ref{Spec}(a). The center conductor with width $S$ is separated from the two ground planes by a distance $W$. The positions of two gaps (of width $G$) in the center conductor define the length of the capacitively coupled resonant structure. The resonators were etched into a 200\,nm thick niobium film (critical temperature $T_\mathrm{c,Nb} \approx 8.3$\,K for our devices) sputtered onto a 430$\,\mu$m thick sapphire substrate, using UV-lithography and a SF$_6$ etching process. Here we show data obtained with two different resonators, \lq S20\rq{} and \lq S40\rq , with fundamental frequencies $f_0 \approx 10$\,GHz and 5\,GHz, respectively. An overview with relevant resonator parameters is found in table \ref{tab1}. 

\begin{table}
		\caption{\label{tab1}Designed values of center conductor width $S$, separation $W$ between inner and outer conductor, gap width $G$, and the fundamental frequency $f_0$ of the resonators.}
\begin{indented}
\item[]\begin{tabular}{@{}lll}
\br
& \lq S20\rq{} &\lq S40\rq{}\\
\mr
		$S$ & 20$\,\mu\meter$ & 40$\,\mu\meter$ \\ 
		$W$ & 8.5$\,\mu\meter$ & 16.8$\,\mu\meter$ \\ 
		$G$ & 30$\,\mu\meter$ & 70$\,\mu\meter$ \\ 
		$f_0$ & 10\,GHz & 5\,GHz \\
\br
\end{tabular}
\end{indented}
\end{table}

Each chip was mounted inside a brass box (see Fig.\ \ref{Spec}(b)), connected to the external circuitry via two 1.85\,mm connectors \cite{ConnectorAnritsu}, and then cooled in a $^4$He glass-cryostat with base temperature of about 1.5\,K. A network analyzer was used to measure the complex transmission coefficient $S_{21}$. The absolute value $|S_{21}|$ versus frequency $f$ for both resonators at a temperature of $T=1.8$\,K is shown in Figs.\ \ref{Spec}(c) and \ref{Spec}(d). In both cases the resonances clearly set themselves apart from the background transmission as equally spaced sharp features up to 50\,GHz (indicated by arrows). 
As commonly found for cryogenic microwave experiments, the measured transmission spectra do not only contain the modes of the coplanar resonator, but also a strongly frequency-dependent background due to unwanted cavity modes in the resonator boxes, standing waves and losses in the transmission lines, all of which become more difficult to handle if a microwave experiment is operated at higher frequencies. 
Full low-temperature calibration of a microwave setup to remove such unwanted influences is demanding \cite{Scheffler2005a,Ranzani2013,Yeh2013}, and therefore we employ a different approach to separate the desired resonances from the background. As described in Ref.\ \cite{Hafner2014}, we measure the background contribution at a temperature of 10\,K, well above $T_\mathrm{c,Nb}$ of the niobium (i.e.\ a spectrum that does not contain any resonant modes of the coplanar device because the Ohmic damping in this metallic configuration suppresses them), and we subtract this spectrum from the transmission spectra below $T_\mathrm{c,Nb}$ as complex quantities. 

\begin{figure}
	\centering
	\includegraphics[width=1.0\linewidth]{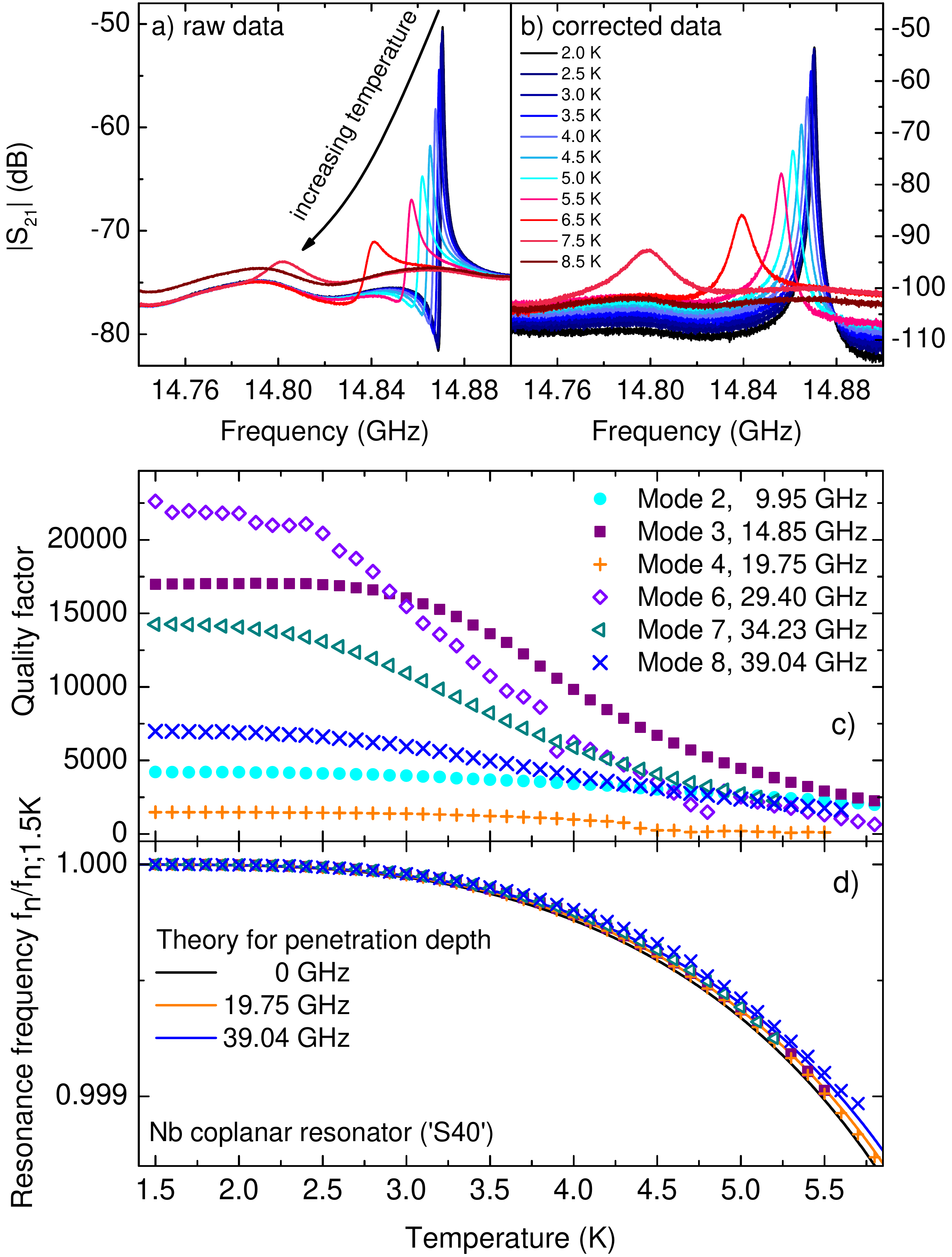}
	\caption{Temperature-dependent performance of the \lq S40\rq{} coplanar resonator. (a) Raw experimental transmission data of the third mode around 15\,GHz for different temperatures. The substantial background contribution can be removed, and (b) a nearly symmetrical lineshape is recovered.
Temperature dependences of (c) quality factors $Q_n$ and (d) resonator frequencies $f_n$ for several modes, with $f_n(T)$ normalized to the respective value $f_{n;1.5\mathrm{K}}$ at the lowest measured temperature. Full lines in (d) are theoretical expectations, following the Mattis-Bardeen formalism. The \lq 0 GHz\rq{} line represents the temperature dependence of the penetration depth for the static case.
}
	\label{Temp}
\end{figure}

\section{Resonator Performance}
As an example, Figs.\ \ref{Temp}(a) and (b) show the temperature dependence of the third mode of the \lq S40\rq{} resonator (fundamental frequency 5\,GHz) before and after subtraction of the background, respectively, where most of the distortions of the measured resonator spectrum are removed by the subtraction. The quality factor $Q_n$ and resonance frequency $f_n$ for each mode (with $n$ index of the mode) were determined by fitting a Lorentzian to the corrected $|S_{21}|^2$ spectra. 
Since the penetration depth and the losses of the superconductor increase upon heating, concomitant  decreases of the resonance frequency and of the quality factor are expected. Fig.\ \ref{Temp}(c) shows the temperature dependence of the quality factor for several modes of this resonator. 
Upon cooling, the quality factor rises rapidly below $T_\mathrm{c,Nb}$ because the electrodynamic losses strongly reduce as the superfluid condensate forms in the superconducting niobium. Below $T\approx 3$~K the quality factor becomes much less temperature dependent, and for our lowest temperatures basically saturates.

Concerning the absolute values of the quality factor, a strong frequency dependence is expected: with increasing frequency, the conductivity of a superconductor strongly decreases and the surface resistance increases \cite{Pracht2013,Hafner2014,Dressel_Buch}, leading to decreasing intrinsic $Q$, and also the capacitive coupling to the feedlines contributes \cite{Goeppl2008}.
Furthermore, the microwave losses of superconductors at lowest temperatures are often governed by residual losses \cite{Halbritter1971} that are not well understood and that for our designs often depend on frequency in an uneven fashion, possibly because the field distributions of the different modes differ in sensitivity concerning the defects e.g.\ in the bends of the center-conductor meander structure. But our data demonstrate that $Q$ above $10^4$ can be achieved for frequencies well beyond 30\,GHz. 

Fig.\ \ref{Temp}(d) shows the temperature dependence of the resonance frequencies $f_n(T)$ for the same resonator, each normalized to the resonance frequency $f_{n;1.5\mathrm{K}}$ at the lowest temperature. The shift in resonance frequency is caused by the change in penetration depth of the niobium, which in turn is governed by the temperature-dependent Cooper-pair density. 
The full lines in Fig.\ \ref{Temp}(d) are calculated using the Mattis-Bardeen formalism \cite{Pracht2013, MattisBardeen1958}: with $T_\mathrm{c,Nb}=8.3$~K, gap-to-$T_\mathrm{c}$ ratio $\Delta_\mathrm{0,Nb}/(k_\mathrm{B} T_\mathrm{c,Nb}) = 1.97$ (consistent with Nb literature \cite{Ashcroft1976,Pronin1998}; with Boltzmann constant $k_\mathrm{B}$), and the frequencies of the experiment as input parameters (and additionally the static case of zero frequency), the temperature dependence of the normalized surface impedance is calculated, which then determines the temperature dependence of the normalized resonator frequencies. 
For $s$-wave superconductors the temperature dependence of the penetration depth should be flat at low $T$ \cite{Oates1991, Hashimoto2009} and frequency independent for $hf\ll 2\Delta$ with optical energy gap $2\Delta$ and Planck constant $h$. In the case of niobium with $2\Delta_0= 2\Delta(T=0) \stackrel{\wedge}{=} $ 720~GHz \cite{Pronin1998}, this condition is met even for our highest accessible frequencies, but only at temperatures well below $T_\mathrm{c,Nb}$. With increasing temperature
, we observe a spreading of the experimental data for the different frequencies that is consistent with the theoretical expectations.

\begin{figure}
	\centering
	\includegraphics[width=1.0\linewidth]{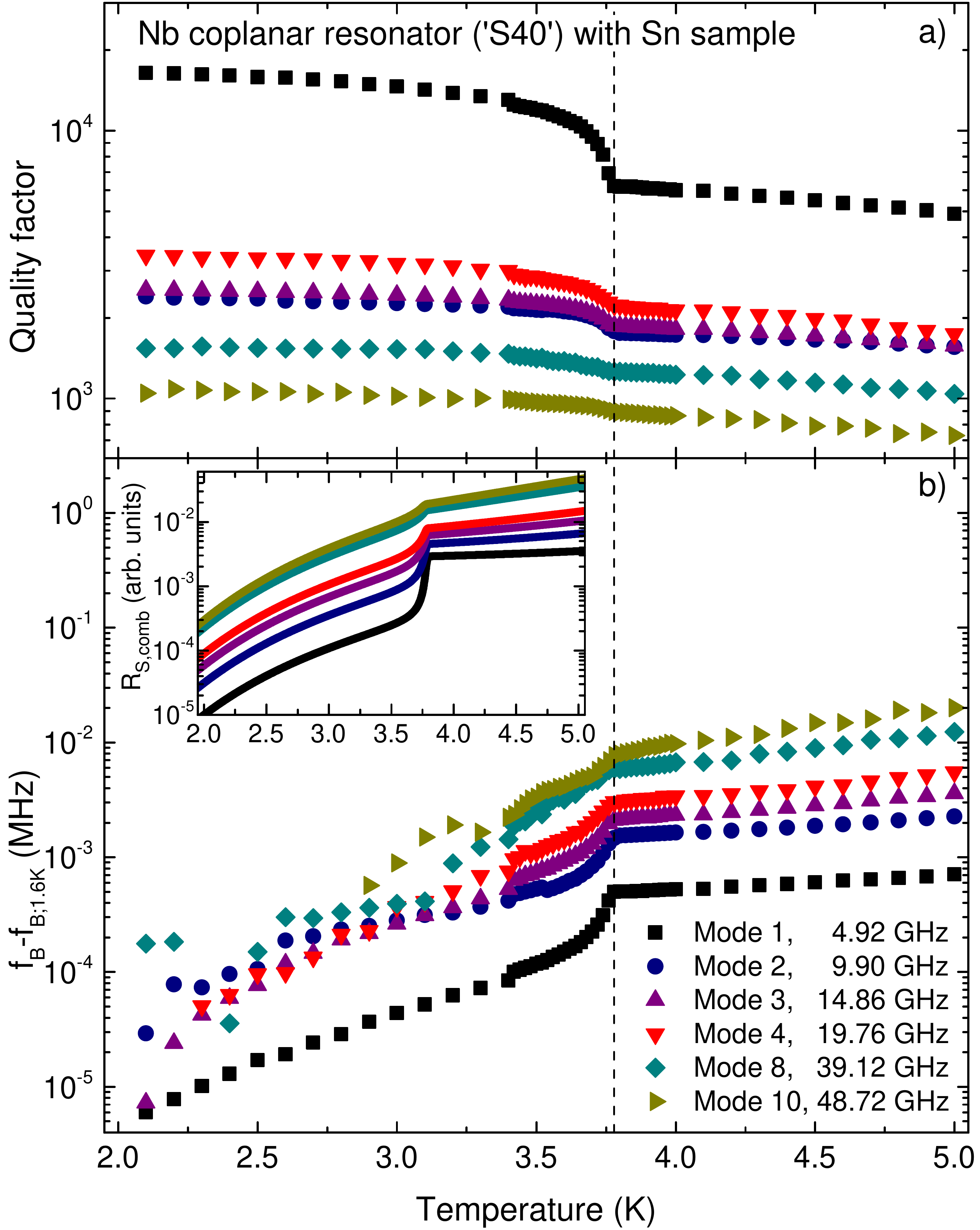}
	\caption{Probing a superconducting tin bulk sample with a coplanar resonator. (a) Temperature dependence of the quality factor of a \lq S40\rq{} resonator with tin sample on top. Six exemplary modes are shown, which display a decrease of the quality factor with increasing temperature. A pronounced transition (dashed line) corresponds to the critical temperature $T_\mathrm{c,Sn}$ of the superconducting tin sample. (b) Temperature dependence of the resonator bandwidth $f_\mathrm{B}$ for the same modes, with bandwidth $f_\mathrm{B;1.6K}$ at lowest temperature subtracted. If the dominant resonator losses are Ohmic, then the bandwidth is proportional to the surface resistance $R_\mathrm{s}$. The inset shows calculated behavior of an effective combined surface resistance $R_\mathrm{s,comb}$ for tin and niobium.}
	\label{Probe}
\end{figure} 

\section{Application: Probing a Bulk Superconductor}
The data presented so far only consider the microwave properties of the bare resonators. 
The high quality factors as well as the only weak temperature dependences of quality factor and resonance frequencies at low temperatures make these resonators a promising spectroscopic tool to investigate the optical properties of different solids in the microwave regime. 
Possible field of application could be the study of samples with intrinsic energy scales below 50\,GHz, such as heavy-fermion systems, where the scattering rate of the quasiparticles shifts into the microwave regime at low temperatures \cite{Scheffler2005c,Scheffler2013} or superconductors with a transition temperature below 1~K, where the superconducting gap $2\Delta$ lies within the microwave regime \cite{Scheffler2015,Thiemann2018a,Thiemann2018b}. 
For demonstration purpose, we investigate the conventional superconductor tin ($T_\mathrm{c,Sn}\approx 3.72$~K \cite{Ashcroft1976}).
A tin foil was placed above a \lq S40\rq{} resonator and thus acts as perturbation to the resonator microwave field.  
A 13\,$\mu$m thick Mylar foil electrically isolated the tin foil from the resonator. 
The temperature dependence of the quality factor of this sample-loaded resonator is shown in Fig.\ \ref{Probe}(a). For temperatures $T>3.8$\,K the quality factor shows a temperature dependence originating from the temperature-dependent response of the niobium resonator, whereas the tin sample is temperature-independent in this regime. 
Upon cooling below $T=3.8$~K an abrupt rise in quality factor is observable originating from the tin sample entering the superconducting phase \cite{Beutel2016, CommentTcSn}. 
This transition is seen up to a frequency of 48.72~GHz, thus proving the interaction of the applied microwaves and the sample at these high frequencies. 
Fig.\ \ref{Probe}(b) shows the temperature dependence of the resonator bandwidth $f_\mathrm{B}$, with the bandwidth $f_\mathrm{B;1.6K}$ at lowest temperature subtracted, for the same resonator modes as in Fig.\ \ref{Probe}(a). 
If the resonator losses are dominated by conductive elements, then their surface resistance $R_\mathrm{s}$ is proportional to $f_\mathrm{B}-f_\mathrm{B;1.6K}$ and can be calculated via cavity perturbation theory \cite{Klein1993}, but requires detailed knowledge of the microwave field geometry with respect to the sample. 
Therefore, we resort to a more qualitative evaluation of these data: optical properties of superconductors are treated within the Mattis-Bardeen theory \cite{MattisBardeen1958, Halbritter}, and our observed temperature dependence in Fig.\ \ref{Probe}(b) qualitatively matches expectations as shown in the inset of Fig.\ \ref{Probe} (b). 
Here the effective overall surface resistance $R_\mathrm{s,comb}$ was calculated assuming a BCS-type temperature dependence of the superconducting energy gaps \cite{TinkhamBook} for tin and niobium (assuming $T_\mathrm{c,Sn} = 3.8$\,K and $\Delta_\mathrm{0,Sn}/(k_\mathrm{B} T_\mathrm{c,Sn}) = 1.76$ for tin and respective values for niobium as above) and a 0.1\% contribution of the Sn sample to the total response of the resonator. A fully quantitative analysis of our data will require more detailed consideration of the microwave field distribution, which is not a plain coplanar structure any more due to the presence of the conducting sample. Furthermore, future refining of the distance between conductive sample and superconducting coplanar resonator will allow improved sensitivity to the high-frequency sample response. Still, the present data clearly demonstrate that the electrodynamics of a sample under study, in this case superconducting tin, can be probed with superconducting coplanar resonators at frequencies up to 50\,GHz.

\section{Conclusions}
Using microwaves in transmission we have shown that superconducting coplanar resonators are well suited for operating frequencies up to 50\,GHz. 
We did not yet optimize the resonators in terms of losses (but rather focused on ease of use). Thus it is not surprising that our observed quality factors are lower than those that were demonstrated with optimized planar superconducting resonators at frequencies below 10~GHz \cite{Megrant2012}, and we expect that substantial increases in quality factors can be achieved with modified designs. But already at this stage we have demonstrated throughout our very wide frequency range quality factors that clearly surpass their non-superconducting counterparts \cite{Javaheri2016}. 
Microwave frequencies well above 20~GHz thus now are available for sensitive spectroscopy studies with compact resonator designs as well as for on-chip quantum optics.

\section*{Acknowledgments}
This work was supported by the Deutsche Forschungsgemeinschaft, including SFB/TRR 21, and by the COST action NANOCOHYBRI (CA16218). 
We thank G.\ Untereiner for experimental support.


\section*{References}

\begin{thebibliography}{99}

\bibitem{Frunzio2005}Frunzio L, Wallraff A, Schuster D, Majer J and Schoelkopf R
2005 \textit{IEEE Tran. Appl. Supercond.} \textbf{15} 860

\bibitem{Hammer2007}Hammer G, Wuensch S, Roesch M, Ilin K, Crocoll E and Siegel M
2007 \textit{Supercond. Sci. Technol.} \textbf{20} S408

\bibitem{Goeppl2008}G\"{o}ppl M, Fragner A, Baur M, Bianchetti R, Filipp S, Fink J M, Leek P J, Puebla G, Steffen L and Wallraff A
2008 \textit{J. Appl. Phys.} \textbf{104} 113904

\bibitem{Javaheri2016}Javaheri Rahim M, Lehleiter T, Bothner D, Krellner C, Koelle D, Kleiner R, Dressel M and Scheffler M
2016 \textit{J. Phys. D: Appl. Phys.} \textbf{49}, 395501

\bibitem{Day2003}Day P K, LeDuc H G, Mazin B A, Vayonakis A and Zmuidzinas J
2003 \textit{Nature} \textbf{425}, 817

\bibitem{Barends2007}Barends R, Baselmans J J A, Hovenier J N, Gao J R, Yates S J C, Klapwijk T M and Hoevers H F C
2007 \textit{IEEE Trans. Appl. Supercond.} \textbf{17}, 263 

\bibitem{Zmuidzinas2012}Zmuidzinas J
2012 \textit{Annu. Rev. Condens. Matter Phys.} \textbf{3}, 169

\bibitem{Barends2014}Barends R \etal
2014 \textit{Nature} \textbf{508}, 500

\bibitem{DiIorio1988}DiIorio M S, Anderson A C and Tsaur B-Y
1988 \textit{Phys. Rev. B.} \textbf{38}, 7019

\bibitem{Langley1991}Langley B W, Anlage S M, Pease R F W and Beasley M R
1991 \textit{Rev. Sci. Instrum.} \textbf{62}, 1801

\bibitem{Porch1995}Porch A, Lancaster M J and Humphreys R G
1995 \textit{IEEE Trans. Microwave Theory Tech.} \textbf{43}, 306

\bibitem{Scheffler2013}Scheffler M \etal 
2013 \textit{Phys. Status Solidi B} \textbf{250}, 439

\bibitem{Hafner2014}Hafner D, Dressel M and Scheffler M
2014 \textit{Rev. Sci. Instrum.} \textbf{85}, 014702

\bibitem{Ghigo2016}Ghigo G, Gerbaldo R, Gozzelino L, Laviano F and Tamegai T
2016 \textit{IEEE Trans. Appl. Supercond.} \textbf{26}, 7300104

\bibitem{Megrant2012}Megrant A \etal
2012 \textit{Appl. Phys. Lett.} \textbf{100}, 113510

\bibitem{Quintana2014}Quintana C M \etal
2014 \textit{Appl. Phys. Lett.} \textbf{105}, 062601

\bibitem{Goetz2016}Goetz J \etal
2016 \textit{J. Appl. Phys.} \textbf{119}, 015304

\bibitem{Plourde2009}Song C, DeFeo M P, Yu K and Plourde B L T
2009 \textit{Appl. Phys. Lett.} \textbf{95}, 232501

\bibitem{Bothner2012}Bothner D \etal
2012 \textit{Appl. Phys. Lett.} \textbf{100}, 012601

\bibitem{Singh2014}Singh V, Schneider B H, Bosman S J, Merkx E P J and Steele G A
2014 \textit{Appl. Phys. Lett.} \textbf{105}, 222601

\bibitem{Ghirri2015}Ghirri A, Bonizzoni C, Gerace D, Sanna S, Cassinese A and Affronte M
2015 \textit{Appl. Phys. Lett.} \textbf{106}, 184101

\bibitem{Thompson2004}Thompson D, Tantot O, Jallageas H, Ponchak G, Tentzeris M and Papapolymerou J
2004 \textit{IEEE Trans. Microw. Theory Tech.} \textbf{52} 1343

\bibitem{Chuang2011}Chuang H-R, Yeh L-K, Kuo P-C, Tsai K-H and Yue H-L
2011 \textit{IEEE Trans. Electron Devices} \textbf{58} 1837

\bibitem{Wallraff2004}Wallraff A, Schuster D I, Blais A, Frunzio L, Huang R-S, Majer J, Kumar S, Girvin S M and Schoelkopf R J
2004 \textit{Nature} \textbf{431}, 162

\bibitem{Gustrau2012}Gustrau F \textit{RF and microwave Engineering} (John Wiley \& Sons, New York, 2012).

\bibitem{Kurpiers2017}Kurpiers P, Walter T, Magnard P, Salathe Y and Wallraff A
2017 \textit{EPJ Quantum Technology} \textbf{4}, 8

\bibitem{Moding}Concerning coaxial cables, many cryogenic microwave setups are already equipped with semirigid cables of 2.2~mm outer diameter (0.086~inch standard) that allow moding-free operation up to 67~GHz.


\bibitem{Pracht2013}Pracht U S \etal
2013 \textit{IEEE Trans. THz Sci. Technol.} \textbf{3} 269

\bibitem{Basov2005}Basov D N and Timusk T
2005 \textit{Rev. Mod. Phys.} \textbf{77} 721

\bibitem{Scheffler2005c}Scheffler M, Dressel M, Jourdan M and Adrian H
2005 \textit{Nature} \textbf{438}, 1135

\bibitem{Maeda2005}Maeda A, Kitano H and Inoue R
2005 \textit{J. Phys.: Condens. Matter} \textbf{17}, R143

\bibitem{Steinberg2008}Steinberg K, Scheffler M and Dressel M,
2008 \textit{Phys. Rev. B} \textbf{77} 214517

\bibitem{Basov2011}Basov D N, Averitt R D, van der Marel D, Dressel M and Haule K
2011 \textit{Rev. Mod. Phys.} \textbf{83} 471

\bibitem{Aslam2015}Aslam N \etal
2015 \textit{Rev. Sci. Instrum.} 86, 064704

\bibitem{Neu2011}Neu E, Steinmetz D, Riedrich-M\"oller J, Gsell S, Fischer M, Schreck M and Becher C
2011 \textit{New J. Phys.} \textbf{13} 025012

\bibitem{Rogers2014}Rogers L J \etal
2014 \textit{Phys. Rev. Lett.} \textbf{113} 263602

\bibitem{Becker2018}Becker J N, Pingault B, Gro\ss D, G\"{u}ndo\u{g}an M, Kukharchyk N, Markham M, Edmonds A, Atat\"{u}re M, Bushev P and Becher C
2018 \textit{Phys. Rev. Lett.} \textbf{120} 053603


\bibitem{Pompeo2007}Pompeo N, Marcon R and Silva E
2007 \textit{J. Supercond. Novel Magn.} \textbf{20} 71

\bibitem{Hering2007}Hering M, Scheffler M, Dressel M and v. L\"{o}hneysen H
2007 \textit{Phys. Rev. B} \textbf{75} 205203

\bibitem{Geiger2016}Geiger D, Pracht U S, Dressel M, Mravlje J, Schneider M, Gegenwart P and Scheffler M
2016 \textit{Phys. Rev. B} \textbf{93} 165131

\bibitem{Tosoratti2000}Tosoratti N, Fastampa R, Giura M, Lenzi V, Sarti S and Silva E
2000 \textit{Int. J. Mod. Phys. B} \textbf{14} 2926

\bibitem{Scheffler2007}Scheffler M, Kilic S and Dressel M
2007 \textit{Rev. Sci. Instrum.} \textbf{78} 086106

\bibitem{Scheffler2010}Scheffler M, Dressel M and Jourdan M
2010 \textit{Eur. Phys. J. B} \textbf{74} 331

\bibitem{Scheffler2005a}Scheffler M and Dressel M
2005 \textit{Rev. Sci. Instrum.} \textbf{76} 074702

\bibitem{Booth1994}Booth J C, Wu D H and Anlage S M
1994 \textit{Rev. Sci. Instrum.} \textbf{65} 2082

\bibitem{Clauss2013}Clauss C, Bothner D, Koelle D, Kleiner R, Bogani L, Scheffler M and Dressel M
2013 \textit{Appl. Phys. Lett.} \textbf{102} 162601

\bibitem{Wiemann2015}Wiemann Y, Simmendinger J, Clauss C, Bogani L, Bothner D, Koelle D, Kleiner R, Dressel M and Scheffler M,
2015 \textit{Appl. Phys. Lett.} \textbf{106} 193505

\bibitem{Scheffler2015}Scheffler M \etal
2015 \textit{Acta IMEKO} \textbf{4} 47

\bibitem{ConnectorAnritsu}Type V102F-R connectors by Anritsu.

\bibitem{Ranzani2013}Ranzani L, Spietz L, Popovic Z and Aumentado J
2013 \textit{Rev. Sci. Instrum.} \textbf{84} 034704

\bibitem{Yeh2013}Yeh J-H and Anlage S M
2013 \textit{Rev. Sci. Instrum.} \textbf{84} 034706

\bibitem{Dressel_Buch}Dressel M and Gr\"uner G 
2002 \textit{Electrodynamics of Solids} (Cambridge: Cambridge University Press)

\bibitem{Halbritter1971}Halbritter J 
1971 \textit{J. Appl. Phys.} \textbf{42} 82

\bibitem{MattisBardeen1958}Mattis D C and Bardeen J
1958 \textit{Phys. Rev.} \textbf{111} 412

\bibitem{Oates1991}Oates D E, Anderson A C, Chin C C, Derov J S, Dresselhaus  G and Dresselhaus M S
1991 \textit{Phys. Rev. B} \textbf{43} 7655

\bibitem{Hashimoto2009}Hashimoto K \etal
2009 \textit{Phys. Rev. Lett.} \textbf{102} 017002

\bibitem{Pronin1998}Pronin A V, Dressel M, Pimenov A, Loidl A, Roshchin I V and Greene L H 
1998 \textit{Phys. Rev. B} \textbf{57} 22

\bibitem{Thiemann2018a}Thiemann M, Beutel M H, Dressel M, Lee-Hone N R, Broun D M, Fillis-Tsirakis E, Boschker H, Mannhart J and Scheffler M
2018 \textit{Phys. Rev. Lett.} \textbf{120} 237002

\bibitem{Thiemann2018b}Thiemann M, Dressel M and Scheffler M
2018 \textit{Phys. Rev. B} \textbf{97} 214516

\bibitem{Ashcroft1976}Ashcroft N W and Mermin N D
1976 \textit{Solid State Physics} (Fort Worth: Saunders College Publishing)

\bibitem{Beutel2016}Beutel M H, Ebensperger N G, Thiemann M, Untereiner G, Fritz V, Javaheri M, N\"{a}gele J, R\"{o}sslhuber R, Dressel M and Scheffler M
2016 \textit{Supercond. Sci. Technol.} \textbf{29} 085011

\bibitem{CommentTcSn}The slight deviation between our measured $T_\mathrm{c,Sn}$ of tin and the literature value might be due to an error in the temperature measurement.

\bibitem{Klein1993}Klein O, Donovan S, Dressel M and Gr\"uner G 
1993 \textit{Int. J. Infrared Millimeter Waves} \textbf{14} 2423

\bibitem{Halbritter}Halbritter J 
1974 \textit{Z. Phys.} \textbf{266} 209

\bibitem{TinkhamBook} Tinkham M 
1996 \textit{Introduction to Superconductivity} 2nd ed. (New York: McGraw-Hill)

\end{thebibliography}
\end{document}